\documentclass[journal]{IEEEtran}
\usepackage{amsmath}
\usepackage{booktabs}
\usepackage{balance}
\usepackage{amssymb} % 用于数学符号
\usepackage{amsfonts} % 用于数学字体

\ifCLASSINFOpdf
\else
   \usepackage[dvips]{graphicx}
\fi
\usepackage{url}

\usepackage{graphicx}

\begin{document}

\title{Distil-DCCRN: A Small-footprint DCCRN Leveraging Feature-based Knowledge Distillation in Speech Enhancement}

% \author{First A. Author, \IEEEmembership{Fellow, IEEE}, Second B. Author, and Third C. Author, Jr., \IEEEmembership{Member, IEEE}
% \thanks{This paragraph of the first footnote will contain the date on which you submitted your paper for review. It will also contain support information, including sponsor and financial support acknowledgment. For example, ``This work was supported in part by the U.S. Department of Commerce under Grant BS123456.'' }
% \thanks{The next few paragraphs should contain the authors' current affiliations, address, and e-mail address. For example, F. A. Author is with the National Institute of Standards and Technology, Boulder, CO 80305 USA (e-mail: author@boulder.nist.gov).}
% \thanks{S. B. Author, Jr., was with Rice University, Houston, TX 77005 USA. He is now with the Department of Physics, Colorado State University, Fort Collins, CO 80523 USA (e-mail: author@lamar.colostate.edu).}}

\author{Runduo Han, Weiming Xu, Zihan Zhang, Mingshuai Liu, Lei Xie, \IEEEmembership{Senior Member, IEEE}

\thanks{Corresponding author: Lei Xie.}
\thanks{Runduo Han, Weiming Xu, Zihan Zhang, Mingshuai Liu and Lei Xie are with the ASLP Lab, School of Computer Science, Northwestern Polytechnical University, Xi’an 710129, China (e-mail: rdhan@mail.nwpu.edu.cn; lxie@nwpu.edu.cn).}}

\markboth{Journal of \LaTeX\ Class Files, Vol. 14, No. 8, August 2015}
{Shell \MakeLowercase{\textit{et al.}}: Bare Demo of IEEEtran.cls for IEEE Journals}
\maketitle

\begin{abstract}

The deep complex convolution recurrent network (DCCRN) achieves excellent speech enhancement performance by utilizing the audio spectrum's complex features. However, it has a large number of model parameters. We propose a smaller model, Distil-DCCRN, which has only 30\% of the parameters compared to the DCCRN. To ensure that the performance of Distil-DCCRN matches that of the DCCRN, we employ the knowledge distillation (KD) method to use a larger teacher model to help train a smaller student model. We design a knowledge distillation (KD) method, integrating attention transfer and kullback-Leibler divergence (AT-KL) to train the student model Distil-DCCRN. Additionally, we use a model with better performance and a more complicated structure, Uformer, as the teacher model. Unlike previous KD approaches that mainly focus on model outputs, our method also leverages the intermediate features from the models’ middle layers, facilitating rich knowledge transfer across different structured models despite variations in layer configurations and discrepancies in the channel and time dimensions of intermediate features. Employing our AT-KL approach, Distil-DCCRN outperforms DCCRN as well as several other competitive models in both PESQ and SI-SNR metrics on the DNS test set and achieves comparable results to DCCRN in DNSMOS. Audio samples are available at \url{https://rdhan3.github.io/Distil_DCCRN_demo/}
\end{abstract}

\begin{IEEEkeywords}
speech enhancement, knowledge distillation %\url{http://www.ieee.org/organizations/pubs/ani_prod/keywrd98.txt}
\end{IEEEkeywords}

\IEEEpeerreviewmaketitle

\vspace{-4pt}

\section{Introduction}

\IEEEPARstart{S}{peech} enhancement (SE) aims to remove the interference and preserve the speech from the noisy mixture. With the advent of data-driven deep learning approaches, SE models can directly learn the relationship between speech and noise from large paired clean-noisy speech datasets and gain a strong ability in noise suppression. More recent approaches prefer using time-frequency (TF) domain models~\cite{Hu2020DCCRNDC, uformer, yin2020phasen}, which have achieved excellent performance by enhancing the noisy speech's real and imaginary components. DCCRN~\cite{Hu2020DCCRNDC} is a popular model for TF domain speech enhancement, and many subsequent works focus on refining this model~\cite{Lv2021DCCRN+,s-dccrn,yang2021dccrn-sub,yeung2022lcdccrn, DCCRN-WPE,liu2023twodccrn,DCCRN-KWS,cdccrn,dccrnvae}.

While the approaches above have achieved remarkable performance, they rely on complicated module designs that improve the model's feature extraction capabilities. However, these designs involve a large number of parameters, making them unsuitable for many parameter-limited small-footprint applications.

In this paper, we aim to design a model with significantly fewer parameters than DCCRN yet achieves comparable performance. Knowledge distillation (KD) methods, which transmit knowledge from a larger teacher model to a smaller student model, are widely used to decrease model size while maintaining comparable performance. Recently, KD has evolved from focusing solely on models outputs~\cite{Hinton2015DistillingTK, cho2019efficacy, mirzadeh2020improved, zhao2022decoupled} to using intermediate features, the outputs from models' middle layers, which hold richer information than model outputs~\cite{Romero2014FitNetsHF,kim2018paraphrasing,tung2019similarity,heo2019comprehensive}. Using a more complicated and high-performing teacher model generally yields better results. However, it also creates significant structural discrepancies and variations in the dimensions of intermediate features between the teacher and student models. To fully distil intermediate features with inconsistent dimensions, several methods have been proposed~\cite{MVAT, ABC-KD, crosslayer, Two-Step}.
However, these approaches are primarily aimed at addressing the variations in the channel dimensions. In the case of audio, the information in the time dimension is also critically important.

TF domain SE models, including DCCRN, require converting audio from the time domain to the TF domain using short-time-fourier-transform (STFT). Due to different requirements of latency time in applications, the teacher and the student models may have different STFT configurations, particularly in hop lengths, which leads to variations in the time dimension of intermediate features. Previous KD methods like attention transfer (AT)~\cite{at} are less effective when both channel and time dimensions are compressed. This complexity makes the features too intricate for the student model to learn directly. To overcome this, we design a new KD method, AT-KL, which combines AT with Kullback-Leibler (KL)~\cite{kullback1997information} divergence. Incorporating KL divergence offers the advantage of aligning the student model's intermediate features distribution with the teacher model, improving the learning process. This AT-KL approach enables the student model to learn the probability distribution of the teacher's compressed features more effectively, significantly enhancing its performance.

By adopting the AT-KL KD method, we propose a small model, Distil-DCCRN, as the student model, which effectively learns the knowledge from a larger and more complicated teacher model, Uformer~\cite{uformer}. Experiments conducted on the DNS dataset show that our proposed Distil-DCCRN achieves DNSMOS~\cite{reddy2022dnsmos} comparable to that of DCCRN, with only 30\% of the parameters of the DCCRN. Remarkably, the PESQ~\cite{pesq} and SI-SNR~\cite{si-snr} of Distil-DCCRN surpass those of DCCRN as well as several other state-of-the-art (SOTA) models.

\begin{figure*}[!htbp]
  \centering
  \includegraphics[height=0.2\textheight, width=0.8\linewidth]{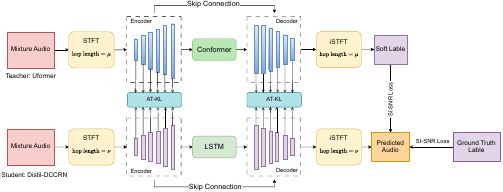}
  \caption{The overall structure of our proposed AT-KL knowledge distillation method. We use Uformer~\cite{uformer} as the teacher model and Distil-DCCRN as the student model. For STFT configuration, their hop lengths are $\mu$ and $\nu$, respectively.}
  \label{sturcture}
\vspace{-8pt}
\end{figure*}

% \begin{figure*}[!htbp]
%   \centering
%   \includegraphics[scale=1.5,height=0.2\textheight]{fig/revision.pdf}
%   \caption{The overall structure of our proposed AT-KL knowledge distillation method. We use Uformer~\cite{uformer} as the teacher model and Distil-DCCRN as the student model. For STFT configuration, their hop lengths are $\mu$ and $\nu$, respectively.}
%   \label{sturcture}
% \vspace{-8pt}
% \end{figure*}

\section{Methods}

\subsection{Overall Structure}

To better validate the effectiveness of the KD method we proposed, AT-KL, we choose Uformer~\cite{uformer} as the teacher model and a dimensionally compressed DCCRN~\cite{Hu2020DCCRNDC} model as the student model, which is named Distil-DCCRN. Their model structures, along with the sizes of the channels and time dimensions of intermediate features, are all different. As shown in Figure \ref{sturcture}, Uformer is a well-designed model that successfully integrates the U-Net structure with attention and performs excellently in speech enhancement. The overall structure of the student model, Distil-DCCRN, is similar to the original DCCRN, using a U-Net structure with long-short-term-memory (LSTM) layers inserted between the encoder and decoder. In Distil-DCCRN, the number of channels in some convolutional and deconvolutional layers of the encoder-decoder and the number of hidden neurons in LSTM layers are decreased compared to the original DCCRN, which is a standard method for decreasing the model's number of parameters.

The AT-KL method implements KD layer by layer between the encoders and decoders of teacher and student models and employs standard distillation between the output of the two models~\cite{Hinton2015DistillingTK}.

\subsection{Attention Transfer}

As mentioned in~\cite{at}, the absolute value of a hidden neuron's activation can indicate the neuron's importance regarding the specific input. Hence, by considering the absolute values of elements in tensor $X$, we can construct an activation map by calculating statistical measures of these values along a particular dimension. Based on this principle, compressing tensor $X$ along a particular dimension can still preserve a significant amount of useful information. Therefore, we use AT to compress a specific dimension of the model's intermediate features, aligning the dimensions of intermediate features between the teacher and student models while retaining a large amount of useful information from the teacher model to guide the student model.

In our study, teacher and student models have misaligned time and channel dimensions. To fully leverage the features of the teacher model in both the time and channel dimensions, we design separate attention transfer mechanisms for each dimension.

\subsubsection{Time Dimension}

In TF domain models, audio signals must first undergo the STFT to transition from the time domain to the TF domain. However, throughout the training phase of the model, the application of varying scenarios necessitates different STFT hop lengths for different models. This leads to a misalignment in the time dimensions of the intermediate features across models. To address this issue, we propose a time dimension AT method that aligns the time dimensions of the teacher and student models while preserving useful information, thereby facilitating KD. 

We define the activation map of the intermediate features in the teacher model as $X^T \in \mathbb{C}^{N \times F \times T}$, where $N$, $F$, and $T$ represent the channel, frequency, and time dimensions, respectively. The time dimension AT takes the activation map $X^T$ as input and compresses the time dimension, which is defined as $F_t$:
\vspace{2pt}
\begin{equation}
F_t: \mathbb{C}^{N \times F \times T} \to \mathbb{C}^{N \times F}
\end{equation}
\vspace{-4pt}
\begin{equation}
F_{t}(X^T) = \sum_{i=1}^{T} |X^T_i|^\lambda
\end{equation}
where $X^T_{i}=X^{T}(; ; i)$ represents activation maps with different time dimensions for the teacher model and the student model. $\lambda$ representing the mapping strength for the time dimension is set to 2, as done in previous works~\cite{at, MVAT}. The absolute value of the activation map can be used as an indicator of importance, as suggested in~\cite{at}. Therefore, we employ this method to compress the time dimension, facilitating knowledge transfer between the teacher and student. We apply an $l2-norm$ to $F_t(X^T)$ to obtain the time dimension compressed activation map $Y^T$: 
$Y^T = \frac{F(X^T)}{||F(X^T)||_2}$.
% \begin{equation}
% Y^T = \frac{F(X^T)}{||F(X^T)||_2}
% \end{equation}

Similarly, we define the activation map of the intermediate features in the student model as $X^S \in \mathbb{C}^{N \times F \times T}$, using the same method to compress the time dimension, which results in the dimension compressed activation map $Y^S$.

\subsubsection{Channel Dimension}

Effective U-Net structured SE models often have more channels in convolution and deconvolution layers in their encoder-decoder modules~\cite{Hu2020DCCRNDC, 
 uformer}. This is because an increase in the number of channels enables the model to extract richer features, which aids in capturing complicated information within the input signal~\cite{s-dccrn}. Each layer of the encoder can be seen as a feature extractor that, with increased depth, can extract higher-level features. However, this inevitably results in a more significant number of model parameters. A standard method to reduce the model's parameter count is by decreasing the number of channels in both the encoder and decoder. Based on this, we propose a channel dimension AT, which aligns the channel dimensions of the teacher and student models while ensuring that valuable information in the channel dimension of the teacher model's intermediate features is preserved.

Similar to the principle of time dimension AT, we compress the channel dimension. It is important to note that we only process activation maps where the teacher and student models have different numbers of channels. We define the activation map of the intermediate features in the teacher model, which has already been compressed in the time dimension, as $Y^T \in \mathbb{C}^{N \times F}$, where $N$ and $F$ represent the channel and frequency dimensions, respectively. We define the channel dimension mapping function $F_c$:
\vspace{2pt}
\begin{equation}
F_n: \mathbb{C}^{N \times F} \to \mathbb{C}^{F}
\end{equation}
\vspace{-4pt}
\begin{equation}
F_{n}(Y^T) = \sum_{j=1}^{N} |Y^T_j|^\lambda
\end{equation}
where $Y^T_{j}=Y(j;)$, $j$ represents the activation maps with different channel dimensions between the teacher and student models. $\lambda$ represents the mapping strength for the channel dimension, set to 2, the same as for the time dimension AT. Then, apply the $l2-norm$ to $F_c(Y^T)$ to obtain the time and channel dimensions compressed activation map $Z^T$: 
$Z^T = \frac{F_c(Y^T)}{||F_c(Y^T)||_2}$.
% \begin{equation}
% Z^T = \frac{F_c(Y^T)}{||F_c(Y^T)||_2}
% \end{equation}

Similarly, we define the time dimension compressed activation map of the intermediate features in the student model as $Y^S \in \mathbb{C}^{N\times F}$ and process the activation maps with different channel dimensions from the teacher model using channel dimension AT, consistent with the method described above, to obtain $Z^S$.

We define the loss function between the compressed activation maps of the student and teacher models as follows:
\begin{equation}
\mathcal{L}_{AT} = \sum_{i=1}^{T} \left\lVert Y^T_i - Y^S_i\right\rVert_{\lambda} + \sum_{j=1}^{N} \left\lVert Z^T_j - Z^S_j\right\rVert_{\lambda}
\end{equation}
% \begin{equation}
% \mathcal{L}_{AT} = \left\lVert \frac{F_c(Y^T)}{\left\lVert F_c(Y^T) \right\rVert_2} - \frac{F_c(Y^S)}{\left\lVert F_c(Y^S) \right\rVert_2} \right\rVert_{\lambda}
% \end{equation}
where $i$ represents the activation map of intermediate features with different time dimensions and 
$j$ represents the activation map of intermediate features with time and channel dimensions differences. $\lambda$ is set to 2, as done in~\cite{MVAT}.

\subsection{AT-KL}
After AT in two dimensions, the intermediate features of the teacher model are not easily learned directly by the student model. Therefore, we make the student model to learn the probability distribution of the teacher model’s intermediate features, a task well-suited for applying KL divergence. To make the student model's prediction of intermediate features as close as possible to the teacher model, thereby reducing the difference in their distributions, we need to minimize the KL divergence. Since KL divergence is a measure of probability, we convert the compressed activation maps of the student and teacher models after AT ($Y^S, Z^S$ and $Y^T, Z^T$) into probability distributions $P$ and $Q$ using the softmax function in advance. Hence, the loss function for KL divergence is defined as:
\begin{equation}
\mathcal{L}_{AT-KL}= \sum_{i=1}^{n} p_i \log\left(\frac{p_i}{q_i}\right)
\end{equation}
$n$ represents the dimension of the frequency axis, and ${p_i}$ and ${q_i}$ represent the probability distributions of the intermediate features at the $i$ th frequency bin for the student model and teacher model, respectively.

\subsection{Knowledge Distillation Method}
When distilling intermediate features, each layer of the encoder-decoder in both the student and teacher models needs to be distilled. For layers with the same channel dimensions, it is only necessary to compress the time dimension using the time dimension AT. However, for layers with different channel dimensions, both time and channel dimension AT must be used to compress both dimensions sequentially. Subsequently, we calculate the KL divergence between the intermediate features activation maps of the teacher and student models, enabling the student model to learn the teacher model's feature distribution further.

In addition to distilling the intermediate features, we also use the teacher model's output as soft labels to distil the student model's output. Since the output of the teacher network is unbounded, it may provide confusing guidance to the student model~\cite{MVAT}. Therefore, we further constrain the training of the student network using ground truth hard labels. We employ the SI-SNR loss~\cite{si-snr} as the loss function for the distilled model outputs, which is represented as follows:
% \begin{equation}
% \mathcal{L}_{SI-SNR}= \alpha\mathcal{L}_{SI-SNR(Hard)} + (1-\alpha)\mathcal{L}_{SI-SNR(Soft)}
% \end{equation}
\begin{align}
L_{\text{SI-SNR}} &= \alpha L_{\text{SI-SNR}}(\text{hard}) \nonumber \\
&\quad + (1 - \alpha) L_{\text{SI-SNR}}(\text{soft})
\end{align}
where $\alpha$ and $1 - \alpha$ is the weight of these loss functions. When $\alpha$ is set to 0.5, the model performs best. 

The loss function of the overall distillation part is defined as follows:
\begin{equation}
\mathcal{L}_{KD}= \beta\mathcal{L}_{SI-SNR} + \gamma\mathcal{L}_{AT} + \eta\mathcal{L}_{AT-KL}
\end{equation}
where $\beta$, $\gamma$ and $\eta$ represent the weight of each loss function. To balance the numerical relationships among the three loss functions, we set $\beta=1$, $\gamma=1$, $\eta=60$ after multiple experiments.

During the model training phase, we first pre-train the teacher model. Then, its parameters are frozen, and the intermediate features and output results are distilled. The student model does not require pre-training.

\section{Experimental Results and Discussion}

\subsection{Datasets}
We train the teacher and student models using the same DNS dataset~\cite{dns_dataset} to compare with other models. The data simulation method is entirely consistent with that in~\cite{Hu2020DCCRNDC}. In the evaluation phase, we use the test set from the DNS dataset for evaluation and select PESQ~\cite{pesq}, SI-SNR~\cite{si-snr}, and DNSMOS~\cite{reddy2022dnsmos} as the evaluation metrics for comparison.

\subsection{Training Setups}
In experiments, we select the Uformer as the teacher model~\cite{uformer}. The STFT window length is 25 ms, with a hop length of 10 ms. The encoder consists of 6 layers of convolutions, with the number of output channels for each convolution layer being [16, 32, 64, 128, 256, 256], and the decoder has the inverse configuration.

We design a student model named Distil-DCCRN, which has a structural resemblance to DCCRN~\cite{Hu2020DCCRNDC} and is also causal. Unlike DCCRN, Distil-DCCRN possesses fewer channels in its encoder and decoder, and the number of hidden neurons in its middle layer LSTM is also decreased. This design strategy results in Distil-DCCRN having total parameters of only 1.1M, which is only 30\% of DCCRN's parameters. Specifically, the STFT window length of Distil-DCCRN is set to 25ms with a hop length of 6.25ms. The encoder consists of six convolutional layers, with the number of output channels for each layer being [16, 32, 64, 64, 128, 256]. The LSTM layers contain 64 hidden neurons.

We choose the winning solution of DNS, DCCRN, as the baseline for comparison, with 3.7M parameters. The output channel numbers for each convolution layer in the encoder are [16, 32, 64, 128, 256, 256], with the LSTM layers having 256 hidden neurons.

Additionally, we compare our models with four other SOTA models with similar parameter counts, namely the Subband Model~\cite{subband}, FullSubnet~\cite{fullsubnet}, GRU-512~\cite{Tan2023CheapNETIL}, and ES-Gabor~\cite{ES-Gabor}. By comparing these well-designed models, further evidence is provided to illustrate the powerful capability of our proposed AT-KL KD method.

During the experiments, we employ the Adam optimizer for training, with an initial learning rate of 0.001, halved whenever the loss increases on two consecutive validation sets. 

\subsection{Experimental Results and Discussion}

As shown in Table \ref{table1}, the teacher model Uformer performs the best, but its parameter count is significantly higher than the other models, reaching 8.8M. 

Compared to the structurally similar DCCRN, the student model Distil-DCCRN, with the assistance of AT-KL KD, achieves higher PESQ and SI-SNR, while the DNSMOS score is also basically the same as DCCRN. Moreover, the number of parameters of Distil-DCCRN is 70\% less than that of DCCRN, and the FLOPs are reduced from 15.7G to 7.4G, only half of DCCRN. Furthermore, DCCRN has become an antiquated design, while numerous contemporary designs enable models to perform similarly to or superior to DCCRN with a decreased number of parameters. However, with the use of AT-KL, Distil-DCCRN matches the performance of SOTA lightweight models, including SubbandModel, FullSubnet, GRU-512, and ES-Gabor, with the only exception being a slight deficit of 0.03 in the WB-PESQ metric compared to ES-Gabor. This underscores the effectiveness of our proposed AT-KL KD approach, which enables simple structured student models to significantly learn the teacher model's capability in extracting speech features.

% \vspace{-6pt}
\begin{table}[!htbp]
\centering
\scriptsize
\setlength{\tabcolsep}{3pt}
\caption{Comparison of the evaluation results of different models on the DNS no reverb test dataset.}
\label{table1}
\vspace{-8pt}
\begin{tabular}{@{}lccccc@{}}
\toprule
Method             & Params (M) & WB-PESQ & NB-PESQ & SI-SNR & DNSMOS \\ \midrule
Uformer (Teacher)   & 8.82      & 3.02    & 3.50    & 19.09  & 3.41   \\
\midrule % 中间的横线
DCCRN~\cite{Hu2020DCCRNDC}             & 3.74      & 2.74    & 3.26    & 17.71  & \textbf{3.22}   \\
SubbandModel~\cite{subband}     & 1.82      & 2.55    & 3.24    & -      & -      \\
FullSubnet~\cite{fullsubnet}        & 5.6       & 2.78    & 3.30    & -      & -      \\
GRU-512~\cite{Tan2023CheapNETIL}           & 3.41      & 2.46    & -       & -      & -      \\
ES-Gabor~\cite{ES-Gabor}          & 1.43      & \textbf{2.83}    & -       & -      & -      \\
% DCCRN-64$^{*}$  & 1.10 & 2.48 & 3.04 & 15.78 & 3.08 \\
% DCCRN-32$^{*}$  & 0.90 & 2.35 & 2.91 & 15.36 & 2.97 \\
Distil-DCCRN  & 1.10      & 2.80    & \textbf{3.31}    & \textbf{17.80}  & 3.21   \\
% Student-0.9(AT-KL)  & \textbf{0.90}      & 2.71    & 3.26    & 17.49  & 3.16   \\ 
\bottomrule
\end{tabular}
\vspace{-12pt}
\end{table}

% \vspace{8pt}

\subsection{Ablation study}

Table \ref{tab:ablation_study} presents the ablation study metrics for Distil-DCCRN, detailing the performance impact of each component on NB-PESQ and SI-SNR. Distil-DCCRN, trained directly without undergoing distillation, achieved an NB-PESQ score of 3.04 and an SI-SNR of 15.78. Distilling only the output results in slightly improved performance, and incorporating AT increased NB-PESQ to 3.23 and SI-SNR to 17.61. Layer-by-layer distillation using KL divergence on the encoder-decoder also yields improvements, with increases of 0.2 and 1.93 in NB-PESQ and SI-SNR, respectively. The best results are obtained by combining AT and KL across all layers, reaching an NB-PESQ of 3.31 and an SI-SNR of 17.81. Additionally, separate distillations using KL divergence only on the encoder or decoder show performance gains, indicating the necessity of utilizing intermediate features from both the encoder and decoder in the KD process.

% \vspace{-8pt}
\begin{table}[!htbp]
\centering
\caption{Ablation study results on the DNS no reverb test dataset.}
\label{tab:ablation_study}
% \vspace{-8pt}
\begin{tabular}{lcc}
\toprule
Method & NB-PESQ & SI-SNR \\
\midrule
Distil-DCCRN (Trained directly) & 3.04 & 15.78 \\
Distil the model output & 3.12 & 16.40 \\
\ \ \ \  + AT & 3.23 & 17.61 \\
\ \ \ \  + KL (Encoder only) & 3.22 & 17.65 \\
\ \ \ \  + KL (Decoder only) & 3.20 & 17.20 \\
\ \ \ \  + KL (All layers) & 3.24 & 17.71 \\
\ \ \ \  + AT + KL & \textbf{3.31} & \textbf{17.81} \\
\bottomrule
\end{tabular}
\end{table}

\vspace{-8pt}
\section{Conclusion }

This study presents an AT-KL KD method that leverages misaligned intermediate features across time and channel dimensions between teacher and student models. Despite decreasing the number of the model's parameters by over 70\%, our AT-KL approach enables the student model, Distil-DCCRN, to achieve DNSMOS comparable to that of DCCRN. Moreover, the PESQ and SI-SNR of Distil-DCCRN not only meet but also exceed those of DCCRN and several other SOTA models. Our experiments on the DNS dataset demonstrate that the AT-KL KD method empowers our smaller, structurally simpler student model to compete with the performance of the latest, well-designed SOTA models. Notably, our proposed AT-KL KD method relies on the versatile U-Net architecture, demonstrating its wide-ranging applicability.

\clearpage
\balance
\bibliographystyle{IEEEtran}
\bibliography{mybib}

\end{document}